\documentclass[aps,prl,preprint,groupedaddress]{revtex4-1}

 \usepackage{graphicx}
 \usepackage{epstopdf}
 \usepackage{color}
 \usepackage[colorlinks=true, urlcolor=navyblue, linkcolor=navyblue, citecolor=navyblue]{hyperref}
\usepackage{lineno}

\def\Dslash{\raise.15ex\hbox{/}\kern-.7em D}
\def\Pslash{\raise.15ex\hbox{/}\kern-.7em P}

\newcommand{\beq}{\begin{equation}}
\newcommand{\enq}{\end{equation}}
\newcommand{\beqa}{\begin{eqnarray}}
\newcommand{\beqast}{\begin{eqnarray*}}
\newcommand{\enqa}{\end{eqnarray}}
\newcommand{\enqast}{\end{eqnarray*}}
\newcommand{\nn}{\nonumber}
\newcommand{\req}[1]{(\ref{#1})}

\newcommand{\mbf}[1]{\mathbf{#1}}

\newcommand{\half}{{\frac{1}{2}}}

\newcommand{\bec}{\begin{center}}
\newcommand{\enc}{\end{center}}
\newcommand{\beqo}{\begin{quote}}
\newcommand{\enqo}{\end{quote}}

\newcommand{\la}{\lambda}

\newcommand{\La}{\Lambda}

\definecolor{navyblue}{rgb}{0,0.08,0.45}
\definecolor{darkred}{rgb}{0.7,0.0,0.0}
\definecolor{darkgreen}{rgb}{0,0.6,0.2}

\begin{document}


\preprint{SLAC--PUB--16257}

\title{Supersymmetry Across the Light and \\Heavy-Light Hadronic Spectrum}

\author{Hans G\"unter Dosch}
\affiliation{Institut f\"ur Theoretische Physik, Philosophenweg 16, 69120 Heidelberg, Germany}
\email[]{h.g.dosch@thphys.uni-heidelberg.de}

\author{Guy F.  de T\'eramond}
\affiliation{Universidad de Costa Rica, 11501 San Pedro de Montes de Oca, Costa Rica}
\email[]{gdt@asterix.crnet.cr}

\author{Stanley J. Brodsky}
\affiliation{SLAC National Accelerator Laboratory, Stanford University, Stanford, California 94309, USA}
\email{sjbth@slac.stanford.edu}

\date{\today}

\begin{abstract}
Relativistic light-front bound-state equations for mesons and baryons can be constructed in the chiral limit  from the supercharges of a superconformal algebra which connect baryon and meson spectra. Quark masses break  the conformal invariance, but the basic underlying supersymmetric mechanism,  which transforms meson and baryon wave functions into each other, still holds and gives remarkable connections across the entire spectrum of light and heavy-light hadrons.  We also briefly examine  the consequences of  extending the supersymmetric relations to double-heavy mesons and baryons.
\end{abstract}

\maketitle

\section{Introduction}

A symmetry relating the baryon and meson spectra was first proposed in~\cite{Miyazawa:1966mfa, Miyazawa:1968zz}, but it was found to be badly broken. The no-go theorem of Coleman and Mandula~\cite{Coleman:1967ad} seemed to put an end to such attempts, since it showed that internal degrees of freedom and space-time symmetries can only be connected to each other in a trivial way.  However,  mainly motivated by work on four-dimensional supersymmetric quantum field theories by Wess and Zumino~\cite{Wess:1974tw}, interest in relating particles by supersymmetry rose sharply in the seventies, since it was shown~\cite{Haag:1974qh} that this symmetry provides  a way to unify space-time and internal symmetries which circumvent the no-go theorem~\cite{Coleman:1967ad}.  Supersymmetry has remained an important underlying principle in particle physics, especially in connection with proposed  extensions of the Standard Model, string theory and grand unification, where supersymmetry was introduced to solve the hierarchy problem~\cite{Dine:2007zp}. Experimental limits require the superpartners of the field quanta to be very massive, and therefore supersymmetry is expected to be broken at the TeV scale.

In 1981, Witten~\cite{Witten:1981nf} introduced supersymmetric quantum mechanics as a model to study nonperturbative supersymmetry breaking, but it was soon realized that this elegant complement to supersymmetric quantum field theory was compelling in its own right~\cite{Cooper:1994eh}. The  simplest form  of supersymmetric quantum mechanics  is generated by two supercharges, the anti-commutator of which is the Hamiltonian of the theory. In fact, as we have shown recently~\cite{Dosch:2015nwa}, the striking empirical similarities of the Regge trajectories of baryons and mesons can be understood as supersymmetric algebraic relations underlying a light-front  (LF) Hamiltonian formulation of confinement in the light quark sector.  This effective theory follows from the clustering properties of the LF Hamiltonian and its holographic embedding in AdS space~\cite{ deTeramond:2008ht, Brodsky:2006uqa, deTeramond:2013it}.  The resulting theory leads to supersymmetric one-dimensional hadronic LF bound-state equations for mesons and baryons,  thus providing a semiclassical approximation to strongly coupled QCD dynamics~\cite{deTeramond:2008ht}.

In our previous papers~\cite{Brodsky:2013ar, deTeramond:2014asa, Dosch:2015nwa}  we have shown how conformal invariance, together with supersymmetric quantum mechanics~\cite{Akulov:1984uh,Fubini:1984hf}, as expressed by holographic LF bound-state equations for light quarks~\cite{deTeramond:2008ht, deTeramond:2013it}, leads to remarkable {\it superconformal}  relations which connect light meson to light baryon spectroscopy~\cite{Brodsky:2014yha}.

As shown in Ref.~\cite{Brodsky:2013ar}, an effective LF Hamiltonian for mesons as bound states of confined light quarks and antiquarks can be derived  for arbitrary spin $J$ based on light-front holography. The confining potential is  determined to have the  form of a harmonic oscillator in the boost-invariant transverse-impact LF variable $\zeta$~\cite{footnote1}.  In fact, the LF confining potential is unique if one requires that the action remain conformally invariant.

In Refs. \cite{deTeramond:2014asa, Dosch:2015nwa}, we showed that the extension of the conformal formalism to a superconformal algebra leads to a remarkable new set of supersymmetric relations for the spectra of light hadrons. In fact, the corresponding construction of the LF Hamiltonian derived from the generalized super charges~\cite{Fubini:1984hf} dictates the form of the LF potential for light mesons and baryons, including constant terms which yield the correct spin dependence and notably a zero-mass pion. This new approach explains the striking similarity of light-quark meson and baryon spectra. A crucial feature of the formalism is that the supermultiplets consist of a meson wave function with internal LF angular momentum  $L_M$ and the corresponding baryon wave function with angular momentum $L_B = L_M - 1$ with the same mass. The $L_M = 0$ meson has no supersymmetric partner.

In this letter we will show that supersymmetric relations between heavy mesons and baryons can also be derived from the supersymmetric algebra even though conformal invariance is explicitly broken by heavy quark masses. We emphasize that the supersymmetric relations which are derived from supersymmetric quantum mechanics are not based on a supersymmetric Lagrangian in which QCD is embedded; instead, they are based on the fact that the supercharges of the supersymmetric algebra relate the wave functions of mesons and baryons in a Hilbert space in which the LF Hamiltonian acts.  The properties of the supercharges predict specific constraints between mesonic and baryonic superpartners in agreement with measurements across the entire hadronic spectrum.

\section{Supersymmetric Quantum Mechanics}

Supersymmetric  quantum mechanics~\cite{Witten:1981nf} can be constructed from the supercharges $Q$ and $Q^\dagger$ with the anticommutation relations  
\beq
 \{Q,Q\} = \{Q^\dagger,Q^\dagger\}=0,
\enq 
and the Hamiltonian 
\beq  {}
H=  \{Q,Q^\dagger\} ,
\enq
which anticommutes with the fermionic generators ${[Q, H]}  = [Q^\dagger, H] = 0$.  Its  minimal realization  in matrix notation is
\beq Q =
\left(\begin{array}{cc}
0&q\\
0&0\\
\end{array}
\right) ,\quad Q^\dagger=\left(\begin{array}{cc}
0&0\\
q^\dagger&0\\
\end{array}
\right) , \label{QQdag}
\enq with 
\beq \label{qdag}
q =-\frac{d}{dx} + W(x), \quad  \quad  q^\dagger = \frac{d}{dx}  + W(x).
\enq  
For the special case $W(x)= \frac{f}{x}$, where $f$ is a dimensionless constant, the resulting Hamiltonian is also invariant under conformal transformations and one can extend the supersymmetric algebra to a  superconformal algebra~\cite{Akulov:1984uh,Fubini:1984hf}.  Furthermore, if one generalizes the supercharge $Q$  to a superposition of fermion generators inside the superconformal algebra~\cite{Fubini:1984hf} by replacing
\beqa  \label{qla}
q &\to& - \frac{d}{dx} + \frac{f}{x} + \la x, \\  \label{qdagla}
q^\dagger &\to&  \frac{d}{dx} + \frac{f}{x} + \la x,
\enqa
in  \req{QQdag}, then the resulting Hamiltonian can be identified with a semiclassical approximation to the QCD LF Hamiltonian of mesons  (M)  and baryons  (B) in the limit of vanishing quark masses~\cite{Dosch:2015nwa}.  Remarkably, the dynamics in this case can be  expressed in terms of a single variable, the  invariant LF transverse coordinate $\zeta$~\cite{footnote1} which is identified with the variable $x$ in Eqs. (\ref{qdag}-\ref{qdagla}).  Additionally, we identify the mass-scale parameters $\la = \la_M = \la_B$  and $f = L_B + \half = L_M - \half $ in the Hamiltonian~\cite{Dosch:2015nwa}, from which follows the crucial relation  $L_M=L_B+1$.

The extension of conformal to superconformal symmetry plays an important role in fixing the effective LF  potentials of both mesons and baryons; thus,  it is interesting  to examine  the role of supersymmetry when conformal symmetry is broken explicitly  by quark masses.  In fact, conformal quantum mechanics was  originally formulated by Witten~\cite{Witten:1981nf} for any form of the superpotential $W(x)$~\req{qdag}: supersymmetry holds if one substitutes the term $\la x$ in \req{qla} and \req{qdagla} by an arbitrary potential $V(x)$. The  resulting Hamiltonian is 
\beq \label{ham} 
H =
\{Q,Q^\dagger\} 
  =     \left(\begin{array}{cc} - \frac{d^2}{d
x^2}+\frac{4(L+1)^2-1}{4x^2}+U_1(x)&\hspace{-1cm}0\\
0&\hspace{-1cm} - \frac{d^2}{d x^2} +\frac{4 L^2-1}{4x^2}+U_2(x)
\end{array}\right), \nn
\enq
$\mbox{with } L=f-\half \mbox{ and }{ U_{1,\,2}(x)} =   \frac{2f}{x}V(x) +V^2(x)\mp V'(x)$.

One can also explicitly break conformal symmetry  without violating supersymmetry by adding  to the Hamiltonian  \req{ham} a multiple of the unit matrix, $\mu^2 \bf I$, where the constant $\mu$  has the dimension of a mass,
\beq  \label{Hnu}
H_\mu =\{Q,Q^\dagger\} + \mu^2  \bf I.
\enq
Interpreting,  as in \cite{Dosch:2015nwa}, the supercharges as transformation operators between the invariant transverse component of the meson  and baryon wave functions~\cite{foot1}, we obtain the same mass relations between mesons and baryons as in the conformal case, and thus their degeneracies remain.  The absolute values of the hadron masses in the heavy quark case, however,  cannot be computed in this framework. This is in contrast to the massless case, where the construction principle  uniquely determines  the LF potential~\cite{deTeramond:2014asa, Dosch:2015nwa}.

Quark masses appear in the LF kinetic energy~\cite{Brodsky:1997de}.  In the case of small quark masses one expects only a small effect on the LF potential. In fact, in~\cite{deTeramond:2014rsa} we have studied the effect of the strange quark mass in the meson sector. It was found that indeed the slopes of the trajectories remain unchanged, indicating that the LF potential is not modified to first order. Small quark masses only affect the longitudinal component of the LF wave function which allows to compute perturbatively  the shift of the squared meson masses. If supersymmetry holds, it then follows  from \req{Hnu} the same shift in the baryonic mass squared with unchanged slope.

\begin{figure} 
\includegraphics[width=16.0cm]{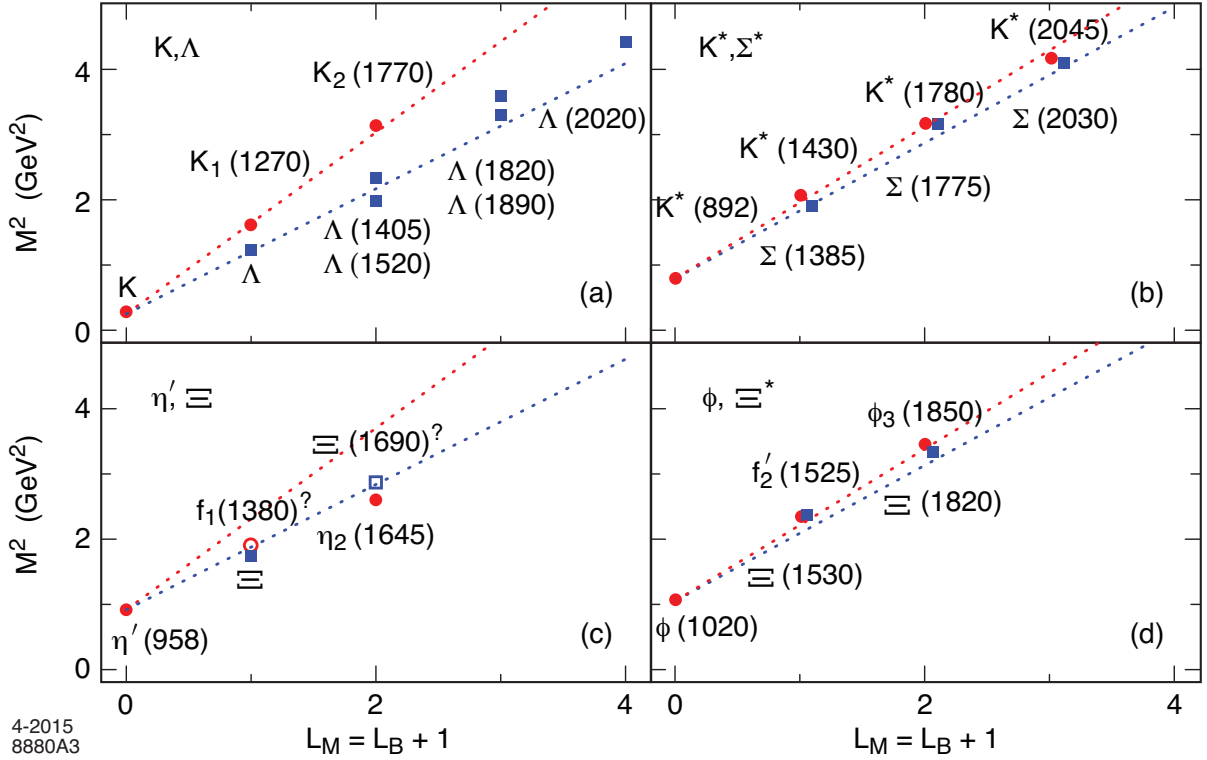}
\caption{ \label{strange}  
Test of supersymmetric relations between  the lowest radial excitations ($n=0$) of strange mesons and  baryons. The dotted lines are the slopes  from the conformal case with  massless quarks~\cite{Brodsky:2014yha}. Hadron masses and assignments are taken from PDG~\cite{Agashe:2014kda}.  Not confirmed states for the $\eta', \Xi$ trajectories are indicated by a white square and white circle.}
\end{figure}

\section{Comparison with Experiment}

We compare in  Fig. \ref{strange} the measured masses for  strange mesons and baryons with the predictions of the supersymmetric model. The squared masses are plotted against $L_M=L_B+1$; mesons and baryons with the same abscissa are predicted to have the same mass. The lightest meson has LF angular momentum zero and therefore can have no supersymmetric baryon partner. The dotted lines in this figure are the trajectories with the slopes taken from the massless case~\cite{Brodsky:2014yha}. Clearly the slopes which describe the Regge trajectories of the light-quark mass hadrons also fit the trajectories of the strange hadrons.  The  supersymmetric relations are remarkably well satisfied for the trajectories of the $K^*$ and $\Sigma^*$ as well as for the  $\phi$ and $\Xi^*$. For the $K$ and $\Lambda$ hadrons there are violations of supersymmetry, similar to those observed for the pion-nucleon system~\cite{Dosch:2015nwa}. These violations  can be traced back to an explicit breaking by the different mass scale values for the mesons and the baryons;  indeed, we have for the K-mesons $\sqrt{\la_K} =\sqrt{\la_\pi} = 0.59$ GeV, whereas 
 for the  $\Lambda$ we have the smaller value $\sqrt{\la_\La} =\sqrt{\la_N} = 0.49$ GeV.  We also include in Fig. \ref{strange}  the $\eta' $ and $\Xi$ trajectories, although the experimental situation is not yet completely settled.

\begin{figure} 
\includegraphics[width=16.0cm]{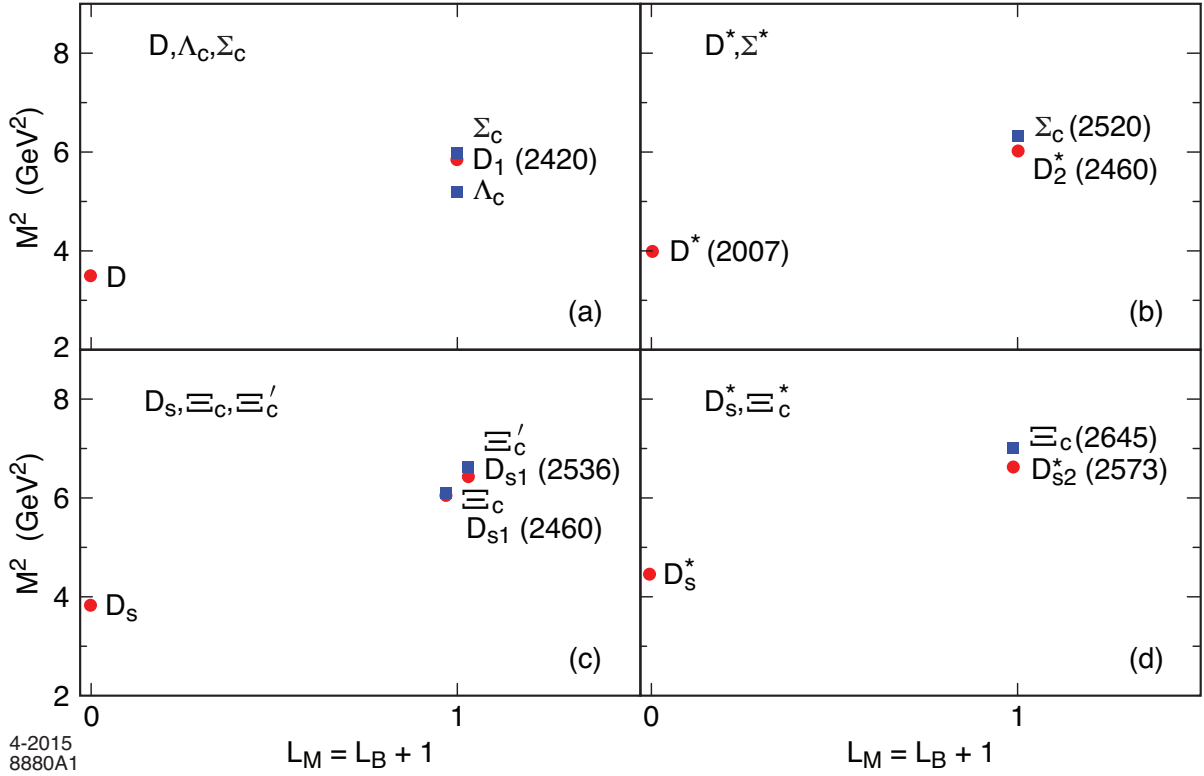}
\caption{\label{heavy1}  Test of supersymmetric relations between mesons and  baryons with charm. Hadron masses and assignments are taken from PDG~\cite{Agashe:2014kda}.}
\end{figure}

\begin{figure} 
\includegraphics[width=16.0cm]{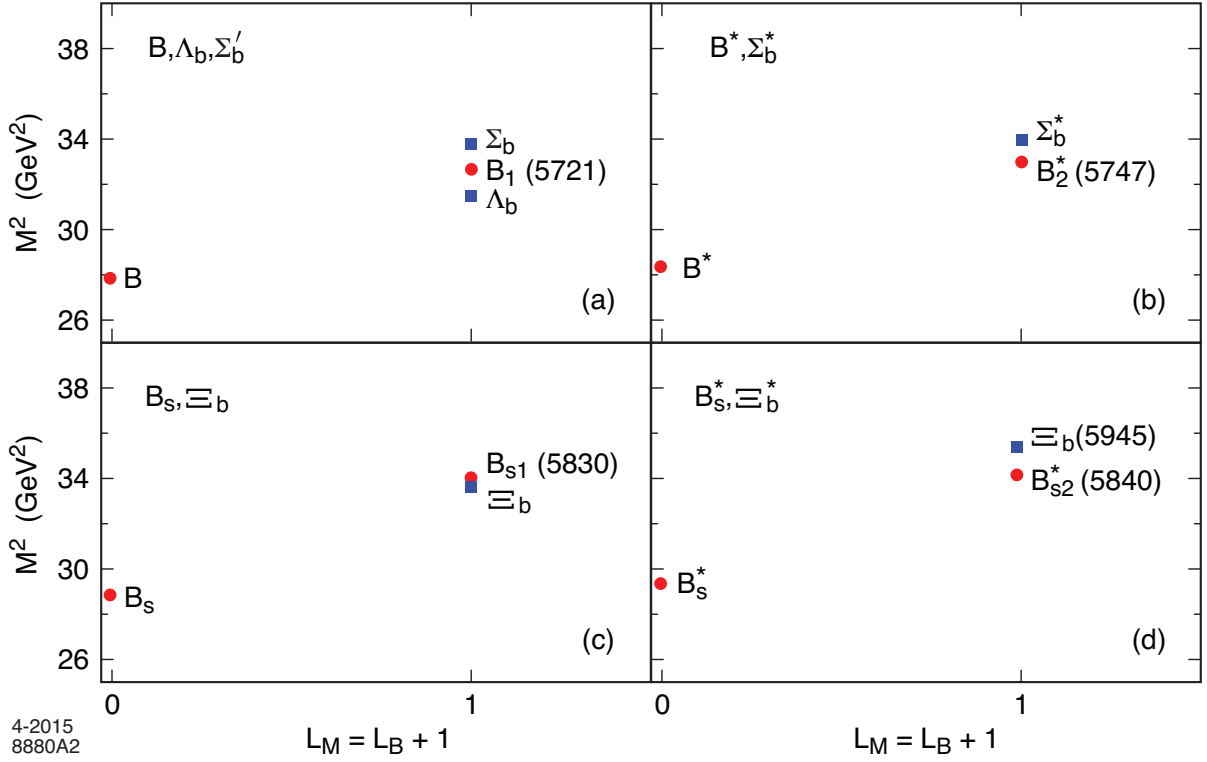}
\caption{\label{heavy2}  Test of supersymmetric relations between mesons and  baryons with beauty. Hadron masses and assignments are taken from PDG~\cite{Agashe:2014kda}.}
\end{figure}

Next we investigate the application of supersymmetry to heavy-light mesons and baryons, namely a meson with one heavy and one light or strange quark, and its corresponding nucleon with one heavy and two light or strange quarks.  Supersymmetry predicts the near degeneracy of mesons with angular momentum $L_M$ and baryons with angular momentum $L_B = L_M -1$.    As can be seen in Figs. \ref{heavy1} and \ref{heavy2} the corresponding mass relations are fulfilled within the expected precision. It is remarkable that the small splitting of the $\Xi_c$ and $\Xi'_c$ is also observed for the corresponding $D_{s1}(2460)$ and $D_{s1}(2536)$ mesons.

One expects large effects from the breaking of conformal symmetry  due to the heavy quark mass. For example, the mass scale $\sqrt{\la}$ need not have a  value similar to that of the light-quark conformal limit. In addition, the confining potential is not required to remain quadratic as prescribed by conformal symmetry. Indeed, the measured difference between the squared mass of the ground state and that of the first orbital excitation is significantly larger than the value obtained from the LF potential between  massless quarks: For the $D$-mesons the discrepancy is a factor of two and  for $B$-mesons  a factor of four  (see also \cite{Branz:2010ub,Gutsche:2012ez}).  The lack of confirmed states does not allow conclusions on the form of the heavy-light LF potential.

Finally, we extend our considerations to double-heavy hadrons; {\it i. e.}, to mesons containing two heavy quarks and their supersymmetric baryon partners containing two heavy and one light quark. This extension should be taken with  care since the kinematical regime for the double-heavy hadronic states is significantly different from the heavy-light systems. With this proviso, we list in Table \ref{doubleheavy-baryons} the double-charm and double-beauty mesons together with the corresponding double-heavy  baryons. Their masses should be degenerate with those of the mesons. These values are higher than the masses of the SELEX double-charm $ccu$ and $ccd$ candidates~\cite{Engelfried:2007at}, but below the predictions from quark models and lattice computations (see Ref.~\cite{Karliner:2014gca} and references quoted therein).

 \begin{table}
 \caption{\label{doubleheavy-baryons} Double-heavy meson states and corresponding double-heavy baryons.}  
\begin{tabular}{cc}
\hline \hline  
Double-heavy meson \quad \quad & \quad \quad Corresponding baryon 
\\ \hline 
$ h_c(1P)(3525) $ & $\Xi_{ccq},\,\half^+ $\\
$ \chi_{c2}(1P)(3556)$ & $\Xi_{ccq}^*,\,\frac{3}{2}^+$\\
$ h_b(1P)(9899)$ & $\Xi_{bbq},\,\half^+ $ \\
$ \chi_{b2}(1P)(9912)$ & $\Xi_{bbq}^*,\,\frac{3}{2}^+$ \\
\hline \hline
\end{tabular} 
 \end{table}

\section{Summary and Conclusions}

In this letter we have described the consequences of  the construction of semiclassical light-front bound-state equations based on supersymmetric quantum mechanics; it relates wavefunctions of mesons and baryons.  The relations are possible since in the light-front holographic approach the baryon is described by the wavefunction of a quark diquark-cluster and the supersymmetric relations reflect the transformation of the antiquark in the meson by a diquark cluster in the baryon~\cite{footnote2}. The foundation of the supersymmetric relations in this approach thus originates in the confinement mechanism of hadrons.  Indeed, when conformal symmetry is restored in the limit of massless quarks, the resulting spectrum of hadronic excitations accounts for essential aspects of hadron spectroscopy~\cite{Brodsky:2013ar,deTeramond:2014asa, Dosch:2015nwa}.
We emphasize that this approach is not based on a supersymmetric quantum field theory.

In the case of  light  quarks, the confining potential is determined uniquely from the underlying conformal invariance~\cite{Brodsky:2013ar,deTeramond:2014asa, Dosch:2015nwa}. In the case of hadrons containing  light and strange quarks, the superpotential has still the same form as for light quarks, but the trajectories undergo a common shift  as can be seen in Fig. \ref{strange}. For systems composed of one heavy and light or strange quarks conformal symmetry is strongly broken but the supersymmetric relations still hold in agreement with experimental observations, as shown in Figs. \ref{heavy1} and \ref{heavy2}. Finally we have extended the supersymmetric relations to systems containing two heavy quarks.

We have shown how supersymmetry, together with light-front holography, leads to new and unexpected connections between mesons and baryons across the hadronic spectrum, thus providing new perspectives for hadron spectroscopy and QCD.   We also note that measurements of additional states in the heavy quark sector will provide important information on the modification of the superpotential due to the explicit breaking of conformal symmetry. This will allow the determination of the light-front confining potential in hadrons for both light and heavy quarks.

\begin{acknowledgments}
We thank the Galileo Galilei Institute for Theoretical Physics for its hospitality and providing an inspiring environment, and the INFN for partial support during the completion of this work. The work of S. J. B. is supported by the Department of Energy Contract No. DE--AC02--76SF00515. 
\end{acknowledgments}


\begin{thebibliography}{99}

 \bibitem{Miyazawa:1966mfa}
  H.~Miyazawa,
  Baryon number changing currents,
 \href{http://ptp.oxfordjournals.org/content/36/6/1266}{Prog.\ Theor.\ Phys.\  {\bf 36},   1266 (1966)}.
 
 
 \bibitem{Miyazawa:1968zz} 
  H.~Miyazawa,
  Spinor currents and symmetries of baryons and mesons,
  \href{http://journals.aps.org/pr/abstract/10.1103/PhysRev.170.1586}{Phys.\ Rev.\  {\bf 170}, 1586 (1968).}
  
   
 \bibitem{Coleman:1967ad} 
  S.~R.~Coleman and J.~Mandula,
  All possible symmetries of the S matrix,
  \href{http://journals.aps.org/pr/abstract/10.1103/PhysRev.159.1251}{Phys.\ Rev.\  {\bf 159}, 1251 (1967)}.
  
  
 \bibitem{Wess:1974tw} 
  J.~Wess and B.~Zumino,
  Supergauge transformations in four dimensions,
  \href{http://www.sciencedirect.com/science/article/pii/0550321374903551}{Nucl.\ Phys.\ B {\bf 70}, 39 (1974)}.
  
  
   \bibitem{Haag:1974qh} 
  R.~Haag, J.~T.~Lopuszanski and M.~Sohnius,
  All possible generators of supersymmetries of the S-matrix,
\href{http://www.sciencedirect.com/science/article/pii/0550321375902795}{Nucl.\ Phys.\  B {\bf 88}, 257 (1975)}.
  
  
  \bibitem{Dine:2007zp} 
  M.~Dine,
  {\it Supersymmetry and string theory: Beyond the standard model}
  (Cambridge University Press, 2007) 515 p.
  
  
  \bibitem{Witten:1981nf}
  E.~Witten,
  Dynamical breaking of supersymmetry,
 \href{http://www.sciencedirect.com/science/article/pii/0550321381900067}{Nucl.\ Phys.\ B {\bf 188}, 513 (1981)}.
 
 
 \bibitem{Cooper:1994eh} 
 For a review see:  F.~Cooper, A.~Khare and U.~Sukhatme,
  Supersymmetry and quantum mechanics,
 \href{http://www.sciencedirect.com/science/article/pii/037015739400080M}{Phys.\ Rep.\  {\bf 251}, 267 (1995)}
  [\href{http://arxiv.org/abs/hep-th/9405029}{\tt arXiv:hep-th/9405029]}.
 
 
 \bibitem{Dosch:2015nwa}
H.~G.~Dosch, G.~F.~de Teramond and S.~J.~Brodsky,
Superconformal baryon-meson symmetry and light-front holographic QCD,
\href{https://journals.aps.org/prd/abstract/17.1103/PhysRevD.91.085016}{Phys.\ Rev.\ D {\bf 91}, 085016 (2015)}
[\href{http://arxiv.org/abs/1501.00959}{\tt arXiv:1501.00959 [hep-th]}].


 \bibitem{deTeramond:2008ht}
  G.~F.~de Teramond and S.~J.~Brodsky,
  Light-front holography: A first approximation to QCD,
  \href{http://prl.aps.org/abstract/PRL/v102/i8/e081601}{ Phys.\ Rev.\ Lett.\  {\bf 102}, 081601 (2009)}
 [\href{http://arXiv.org/abs/0809.4899}{\tt arXiv:0809.4899 [hep-ph]}].
 
 
 \bibitem{Brodsky:2006uqa}
  S.~J.~Brodsky and G.~F.~de Teramond,
  Hadronic spectra and light-front wave functions in holographic QCD,
  \href{http://prl.aps.org/abstract/PRL/v96/i20/e201601}{ Phys.\ Rev.\ Lett.\  {\bf 96}, 201601 (2006)}
  [\href{http://arXiv.org/abs/hep-ph/0602252}{\tt arXiv:hep-ph/0602252}].
 
 
 \bibitem{deTeramond:2013it}
 G.~F.~de Teramond, H.~G.~Dosch and S.~J.~Brodsky,
 Kinematical and dynamical aspects of higher-spin bound-state equations in holographic QCD,
 \href{http://prd.aps.org/abstract/PRD/v87/i7/e075005}{Phys.\ Rev.\ D {\bf 87}, 075005 (2013)}
 [\href{http://arxiv.org/abs/arXiv:1301.1651}{\tt arXiv:1301.1651 [hep-ph]}].  
 
  
 \bibitem{Brodsky:2013ar} 
  S.~J.~Brodsky, G.~F.~de Teramond and H.~G.~Dosch,
  Threefold complementary approach to holographic QCD,
  \href{http://www.sciencedirect.com/science/article/pii/S0370269313010198}{Phys.\ Lett. \ B {\bf 729}, 3 (2014)}
 [\href{http://arxiv.org/abs/arXiv:1302.4105}{\tt arXiv:1302.4105 [hep-th]}].
 
 
  \bibitem{deTeramond:2014asa}
  G.~F.~de Teramond, H.~G.~Dosch and S.~J.~Brodsky,
 Baryon spectrum from superconformal quantum mechanics and its light-front holographic embedding,
\href{http://journals.aps.org/prd/abstract/17.1103/PhysRevD.91.045040}{Phys.\ Rev.\ D {\bf 91}, 045040 (2015)}
 [\href{http://arxiv.org/abs/arXiv:1411.5243}{\tt arXiv:1411.5243 [hep-ph]}].

 
 \bibitem{Akulov:1984uh}
  V.~P.~Akulov and A.~I.~Pashnev,
  Quantum superconformal model in (1,2) space,
  \href{http://link.springer.com/article/17.1007%2FBF01086252}{Theor.\ Math.\ Phys.\  {\bf 56}, 862 (1983)}
  [Teor.\ Mat.\ Fiz.\  {\bf 56}, 344 (1983)].


 \bibitem{Fubini:1984hf}
  S.~Fubini and E.~Rabinovici,
  Superconformal quantum mechanics,
 \href{http://www.sciencedirect.com/science/article/pii/055032138490422X}{Nucl.\ Phys.\ B {\bf 245}, 17 (1984)}.
 
  
  \bibitem{Brodsky:2014yha}
  S.~J.~Brodsky, G.~F.~de Teramond, H.~G.~Dosch and J.~Erlich,
 Light-front holographic QCD and emerging confinement,
 Phys. Rep. (to be published)
  [\href{http://arxiv.org/abs/arXiv:1407.8131}{\tt arXiv:1407.8131 [hep-ph]}].
 
    
 \bibitem{footnote1}
 For $n$ partons the invariant LF variable $\zeta$ is the $u$-weighted definition of the transverse impact variable of the $n-1$ spectator system~\cite{Brodsky:2006uqa}: $\zeta = \sqrt{\frac{u}{1-u}} \big\vert \sum_{j=1}^{n-1} u_j \mbf{b}_{\perp j} \big\vert$, where $u_j$ and $u$ are the longitudinal momentum fractions of quark $j$ in the cluster and of the active quark, respectively.  For a two-parton bound state $\zeta = \sqrt{u(1-u)} \vert \mbf{b}_\perp \vert$.  For a baryon, the LF cluster decomposition corresponds to a quark diquark-cluster decomposition.  In LF holographic QCD the variable $\zeta$ is identified with the bulk variable $z$~\cite{ deTeramond:2008ht, Brodsky:2014yha}.
  
  
 \bibitem{foot1}
 In fact, the lower component is identified with the leading twist (positive chirality) component of the nucleon wavefunction~\cite{Dosch:2015nwa}.
 
 
  \bibitem{Brodsky:1997de}
  S.~J.~Brodsky, H.~C.~Pauli and S.~S.~Pinsky,
  Quantum chromodynamics and other field theories on the light cone,
  \href{http://www.sciencedirect.com/science/article/pii/S0370157397000896}{Phys.\ Rept.\  {\bf 301}, 299 (1998)}
  [\href{http://arXiv.org/abs/hep-ph/9705477}{\tt arXiv:hep-ph/9705477}].
  
 
 \bibitem{deTeramond:2014rsa} 
  G.~F.~de Teramond, S.~J.~Brodsky and H.~G.~Dosch,
  Light-front holography in QCD and hadronic physics,
\href{http://arxiv.org/abs/arXiv:1405.2451}{\tt arXiv:1405.2451 [hep-ph]}.


 \bibitem{Agashe:2014kda}
  K.~A.~Olive {\it et al.}  (Particle Data Group Collaboration),
  Review of particle physics,
\href{http://iopscience.iop.org/1674-1137/38/9/090001/}{Chin.\
Phys.\ C {\bf 38}, 090001 (2014).}


\bibitem{Branz:2010ub}
  T.~Branz, T.~Gutsche, V.~E.~Lyubovitskij, I.~Schmidt and A.~Vega,
  Light and heavy mesons in a soft-wall holographic approach,
   \href{http://prd.aps.org/abstract/PRD/v82/i7/e074022}{ Phys.\ Rev.\  D {\bf 82}, 074022 (2010)}
 [\href{http://arXiv.org/abs/10016.0268}{\tt arXiv:10016.0268 [hep-ph]}].


 \bibitem{Gutsche:2012ez}
  T.~Gutsche, V.~E.~Lyubovitskij, I.~Schmidt and A.~Vega,
Chiral symmetry breaking and meson wave functions in soft-wall AdS/QCD,
  \href{http://prd.aps.org/abstract/PRD/v87/i5/e056001}{Phys.\ Rev.\ D {\bf 87},  056001 (2013)}
  [\href{http://arxiv.org/abs/arXiv:1212.5196}{\tt arXiv:1212.5196 [hep-ph]}].
  
  
 \bibitem{Engelfried:2007at} 
  J.~Engelfried [SELEX Collaboration],
  SELEX: Recent progress in the analysis of charm-strange and double-charm baryons,
  [\href{http://arxiv.org/abs/hep-ex/0702001}{\tt  arXiv:hep-ex/0702001}].  
 
  
\bibitem{Karliner:2014gca} 
  See M.~Karliner and J.~L.~Rosner,
 Baryons with two heavy quarks: Masses, production, decays, and detection,
\href{http://journals.aps.org/prd/abstract/10.1103/PhysRevD.90.094007}{Phys.\ Rev.\ D {\bf 90}, 094007 (2014)}
  [\href{http://arxiv.org/abs/1408.5877}{\tt arXiv:1408.5877 [hep-ph]}].  
  
  
 \bibitem{footnote2}
 Specific quark-diquark models relating mesons and baryons are given in Refs.~\cite{Catto:1984wi, Catto:2008xx, Lichtenberg:1999sc, Mussa:2015}

 
 \bibitem{Catto:1984wi}
  S.~Catto and F.~Gursey,
  Algebraic treatment of effective supersymmetry,
  \href{http://link.springer.com/article/10.1007%2FBF02902548}{Nuovo Cim.\ A {\bf 86}, 201  (1985).}
  
  
 \bibitem{Catto:2008xx}
  S.~Catto,
  Miyazawa supersymmetry,
 \href{http://scitation.aip.org/content/aip/proceeding/aipcp/10.1063/1.2932297}{AIP \ Conf. \ Proc. {\bf 1011}, 253 (2008)}.
 

\bibitem{Lichtenberg:1999sc}
  D.~B.~Lichtenberg,
  Whither hadron supersymmetry?,
\href{http://arxiv.org/abs/hep-ph/9912280}{\tt
arXiv:hep-ph/9912280}.
 
  
\bibitem{Mussa:2015}
R. Mussa, 
Hadron spectroscopy at the $e^+ e^-$ colliders/B factories, Presented at the  7$^{th}$ International Conference on Quark and Nuclear Physics (QNP 2015), Valparaiso, 2-6 March 2015,  \href{https://indico.cern.ch/event/304663/contribution/55/material/slides/0.pdf}{\tt https://indico.cern.ch/event/304663/contribution/55/material/slides/0.pdf}.
  

\end{thebibliography}
\end{document}